\documentclass{PoS}

\newcommand{\bea}{\begin{eqnarray}}
\newcommand{\eea}{\end{eqnarray}}
\newcommand{\beq}{\begin{equation}}
\newcommand{\eeq}{\end{equation}}

\def \thl {{\theta_l}}
\def \thK {{\theta_{K^*}}}

\title{Theoretical Status of Rare B Decays with Muons }

\ShortTitle{Theoretical Status of Rare B Decays with Muons }

        \author{Christoph Bobeth\\
        IPHC, Universit{\'e} de Strasbourg, CNRS/IN2P3, F-67037, Strasbourg, France}

\author{\speaker{Gudrun Hiller}\thanks{Preprint DO-TH 09/18} $\,$ and Giorgi Piranishvili%
   \\
        Institut f\"ur Physik, Technische Universit\"at Dortmund,
  D-44221 Dortmund, Germany}
        
\abstract{
Rare $B$ decays allow to investigate fundamental
interactions regarding their flavor, chiral, Dirac and CP properties.
In anticipation of the large data samples of exclusive $B$ decays into muons from the forthcoming LHC experiments, in particular LHCb, as well as possible super flavor factories, we review the theoretical status and outline future opportunities to explore the borders of the Standard Model and beyond.}

\FullConference{12th International Conference on B-Physics at Hadron Machines \\
                 September 7-11, 2009\\
                 Heidelberg, Germany.}

\begin{document}

\section{Introduction}

The quest for physics Beyond the Standard Model (BSM) is of highest priority at current 
and future flavor facilities.
We focus here on exclusive semileptonic decays induced by
$ b\to s l^+l^-$ transitions to test the Standard Model (SM) and probe BSM physics.  The decays into muons $l=\mu$ are especially well suited for investigations at
hadron colliders, such as the Tevatron and most important in terms of high luminosity, the Large Hadron Collider (LHC). 
The observables presented are also of  relevance to super flavor $e^+ e^-$ facilities.

\section{A brief $b \to s \mu^+ \mu^-$ primer}

We review briefly the current experimental situation for rare $b \to s l^+l^-$ processes, 
recap the theory framework how to extract BSM couplings from $\Delta B=1$ observables, and comment on cuts.

\subsection{The experimental situation}

Important modes for hadron colliders are the exclusive decays
$B \to K^{(*)} \mu^+ \mu^-$, $B_s \to \Phi \mu^+ \mu^-$ and 
$\Lambda_b \to \Lambda \mu^+ \mu^-$ with SM branching ratios of 
the order ${\cal{B}}_{SM} \sim 10^{-7}-10^{-6}$. 
The decays $B \to 
Kø\mu^+ \mu^-$  and  $B \to K^* \mu^+ \mu^-$ have been observed at the $B$ factory experiments Belle and BaBar \cite{Barberio:2008fa}
\bea
{\cal{B}}(B  \to K  \mu^+ \mu^-) =(0.48 \pm 0.06) \cdot 10^{-6} , ~~~~~~~~
{\cal{B}}(B  \to K^*  \mu^+ \mu^-) = (1.15^{+0.16}_{-0.15}) \cdot 10^{-6} , 
\eea
in agreement with the SM.  The Tevatron is close to seeing ${\cal B}(B_s \to \Phi \mu^+ \mu^-)$
\cite{beautymorello}.
Experimental investigations of more involved (and also more BSM diagnostical) observables 
such as dilepton mass spectra, lepton angle distributions and dimuon to dielectron ratios are 
currently underway \cite{:2008ju,Aubert:2008ps,:2009zv}, see Sec~\ref{sec:earlydata} and \ref{sec:full}.
The purely leptonic decay
$B_s \to \mu^+ \mu^-$ is very rare in the SM, ${\cal{B}}_{SM} \simeq 3 \cdot 10^{-9}$, but can be
significantly enhanced in models that circumvent the lepton mass suppression
present in the SM. Currently a bound only exists for its branching ratio 
${\cal{B}}(B_s \to \mu^+ \mu^-)  < 3.6 \cdot 10^{-8}$ @90\% C.L.~\cite{CDF9892}.
Frequently averages over $e^+ e^-$ and $\mu^+ \mu^-$ final states are quoted
for the semileptonic $b \to s l^+l^-$ observables. Indeed the SM predictions agree up to very small corrections after
appropriate cuts in the dilepton mass $q^2$ have been taken into account \cite{Hiller:2003js}.
However, the averaging washes out  possible lepton flavor
non-universal effects, such as  from Higgs exchanges, leptoquarks or R-parity violation,
and therefore applies only to a restricted set of  BSM models.
So far inclusive decays into dimuons have not been observed (at $\geq 5 \sigma$) yet, however, 
the lepton average $(l=e,\mu)$ is observed
${\cal{B}}(B  \to X_s  l^+ l^-) = (3.66^{+0.76}_{-0.77}) \cdot 10^{-6}$  for $q^2 > 0.04 \, \mbox{GeV}^2$ \cite{Barberio:2008fa}.

\subsection{ The effective theory}

\begin{figure} 
\begin{center}
\includegraphics[width=.6\textwidth]{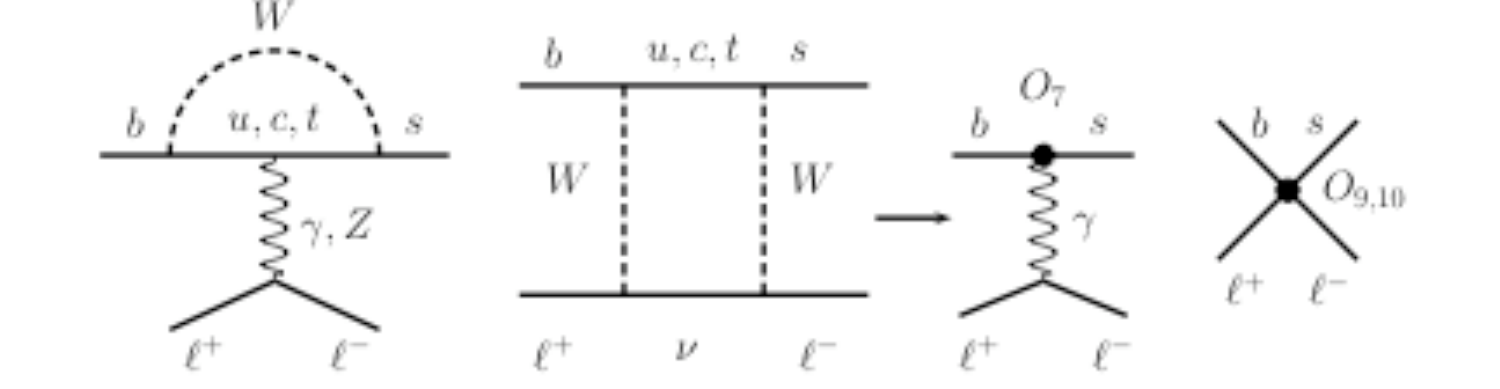} 
\end{center}
\caption{Lowest order Feynman diagrams in the SM contributing to $b \to s l^+ l^-$
in the full (left) and effective theory (right).} 
\label{fig:SMdiags} 
\end{figure}

Our aim is to test the SM and probe BSM with quantum loop effects. The framework used is a generalized Fermi theory of electroweak interactions valid for external momenta
much below the scale where electroweak interactions,  {\it e.g.}, for $B$ physics $m_b^2 \ll m_W^2$,  are induced
\begin{eqnarray} \label{eq:heff}
{\cal{H}}_{\rm eff}= -4 \frac{G_F}{ \sqrt{2}}  V_{tb} V_{ts}^* 
\sum_i C_i(\mu) O_i(\mu) .
\end{eqnarray}
A picture of matching the full theory, ${\cal{L}}$, onto the effective one, Eq.~(\ref{eq:heff}),
is given in Fig.~\ref{fig:SMdiags} for the SM.
New Physics (NP) can appear  in  the Wilson coefficients $C_i=C_i^{\rm SM} + C_i^{\rm NP}$ or in new operators $O_i$.
The  effective theory framework allows for a 
model-independent analysis to determine the $C_i$ from multi-observables/multi-processes
\cite{Ali:1994bf}. 
Couplings for $b \to s l^+ l^-$ are given in Table \ref{tab:Ci}.
\begin{table}[ht]
\renewcommand{\arraystretch}{1.0}
         \begin{center}
         \begin{tabular}{c|c|c|c} 
Wilson coefficient & description &  SM  & enhancement in models
\\ \hline 
$C_{1,2}$ & charged current & YES  & \\
\hline
$C_{3 ,..,6}$ & QCD penguins & YES  &  SUSY \\
$C_{7,8}$ & $\gamma,g$-dipole  & YES  &  SUSY, large $\tan \beta$\\
$C_{9,10}$ & (axial-)vector & YES  &  SUSY \\
\hline
$C_{S,P}$ & (pseudo-)scalar   & $\sim m_l m_b/m_W^2$ 
 &  SUSY, large $\tan \beta$, R-parity viol.\\
$C^\prime_{S,P}$ & (pseudo-)scalar  flipped &  $\sim m_l m_s/m_W^2$  &  SUSY, R-parity viol.\\
$C^\prime_{3 ,..,6}$ & QCD peng. flipped & $\sim m_s/m_b$ &  SUSY \\
$C^\prime_{7,8}$ & $\gamma,g$-dipole flipped & $\sim m_s/m_b$ &  SUSY, esp. large $\tan \beta$\\
$C^\prime_{9,10}$ & (axial-)vector flipped& $\sim m_s/m_b$  &  SUSY \\
$C_{T,T5}$ & tensor  & negligible &  leptoquarks\\
         \end{tabular}
         \end{center}
         \caption{Effective couplings for $b \to s l^+ l^-$ and appearance in various models, see, {\it e.g.},
         \cite{Bobeth:2007dw}. The flipped operators $O^\prime_i$ are obtained from the $O_i$ by interchanging the  chiralities $L \leftrightarrow R$.}
         \label{tab:Ci}
\end{table}
Within the SM the decays $b \to s l^+l^-$ are well described by ten operators with real coefficients.
In general, the number of operators consistent with gauge and Poincare invariance is more than twice as large. The Wilson coefficients can also carry CP phases.
In addition $b \to s l^+l^-$ transitions depend in general on the lepton flavor, 
{\it i.e.}, $C_i  \to C_i^l$, and
 even lepton flavor violation maybe considered with the additional operators 
 $O_i^{l, l^\prime}\sim \bar s \Gamma b \bar l \Gamma^\prime l^\prime$.
Since the resulting number of orthogonal observables required for an extraction of the complete
effective operator structure in ${\cal{H}}_{\rm eff}$ is  very large, constrained model frameworks are useful, such as minimal flavor violation, where $C_i^\prime/C_i \sim m_s/m_b$.

\subsection{Dilepton invariant mass cuts   \label{sec:cuts}}

Kinematical cuts are vital to reduce the background from intermediate charmonia via $b \to s \Psi^{(n)}( \to l^+ l^-) \to s   l^+ l^-$. One distinguishes the low $q^2$ region
 with $q^2 < m_{J/\Psi}^2$ and the high $q^2$ region with $q^2 >m_{\Psi^\prime}^2$.
Another important reason to use cuts is that there is no single rigorous theory framework available for exclusive  $b \to s l^+l^-$ decays in the whole kinematical region.
Only $q^2$-binned data allow for a systematic comparison with theory.
Theoretically preferred is the low dilepton mass below the $J/\Psi$, where many
works exist,  {\it e.g.}, \cite{Beneke:2004dp}. The
high $q^2$ region is calculable with an $1/\sqrt{q^2},1/m_b$ expansion \cite{Grinstein:2004vb}.
The whole $q^2$ region tests the SM, and different regions are sensitive to different NP couplings and models.

\subsection{Forward-backward asymmetry and early data \label{sec:earlydata}}

\begin{figure} 
\begin{center}
\includegraphics[width=.3\textwidth]{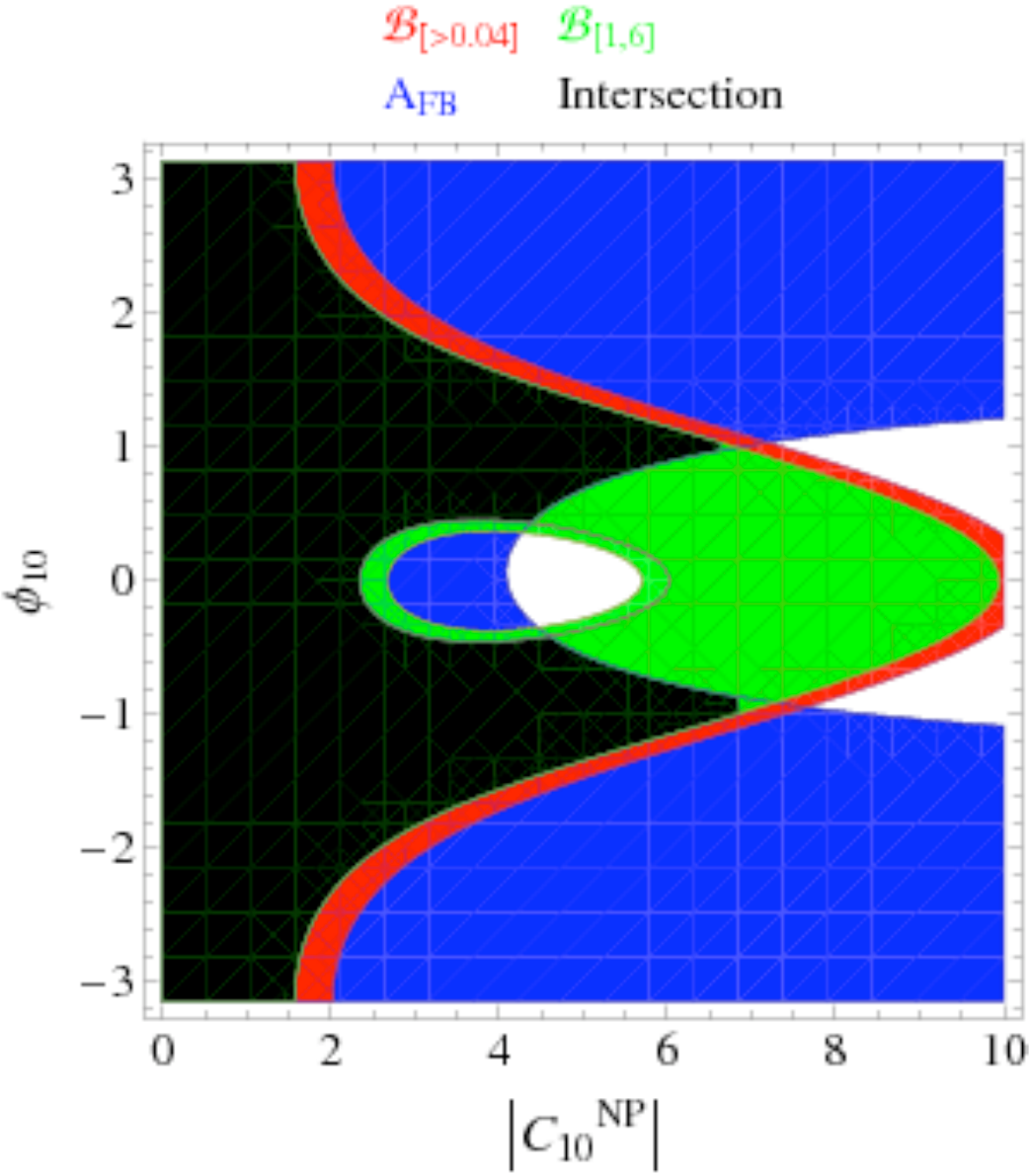} 
\end{center}
\caption{Constraints on the magnitude and CP phase of the NP contribution to $C_{10}$ using different, complementary measurements.The black area is allowed
by ${\cal{B}}$ (green, red) and $A_{\rm FB}$ (blue) constraints \cite{Bobeth:2008ij}.}
\label{fig:C10bounds} 
\end{figure}

Consider the forward-backward asymmetry $A_{\rm FB}$ in $B \to K^* l^+l^-$ decays.
The low $q^2$ region is sensitive to ${\rm sign} \,C_7$, whereas the
high $q^2$ region probes the 4-Fermi  operators, $ {\rm sign} \,C_9^* C_{10}$.
Recent data strongly favor the sign of $A_{\rm FB}$  in the
high $q^2$ region to be SM-like:
 $A_{\rm FB}^{high \, q^2} = 0.76^{+0.52}_{-0.32} \pm 0.07$ (BaBar) \cite{:2008ju}, 
$A_{\rm FB}^{high \, q^2>16 \mbox{\scriptsize GeV}^2} = 0.66^{+0.11}_{-0.16} \pm 0.04$ 
(Belle) \cite{:2009zv}.
Already fixing ${\rm sign} \, A_{\rm FB}^{high \, q^2}>0$ yields useful constraints which are orthogonal to the ones
from the branching ratio measurements  \cite{Bobeth:2008ij}, see Fig.~\ref{fig:C10bounds}.
Here the allowed region of the phase and magnitude of the NP contribution to $C_{10}$
is shown.  Note that
$O_{10} \sim \bar s_L \gamma_\mu b_L \bar l \gamma^\mu \gamma_5 l$ captures the effect of 
non-standard $bZs$-penguins, while their contribution to $O_9\sim \bar s_L \gamma_\mu b_L \bar l \gamma^\mu l$ is suppressed by $(1-4 \sin^2 \theta_W) \ll1$ \cite{Buchalla:2000sk}.

\section{The future: Angular analysis \label{sec:full}}

After observing  the $B \to K^{(*)} l^+ l^-$ decays and early measurements of their
asymmetries \cite{:2008ju,Aubert:2008ps,:2009zv} one can use these modes for detailed investigations of the $b \to s$ transitions and the structure of ${\cal{H}}_{\rm eff}$, Eq.~(\ref{eq:heff}).
A multitude of complementary observables can be obtained  from 
the angular distributions in $B \to K^* (\to K \pi) l^+l^-$     
\cite{Bobeth:2008ij},\cite{Kruger:1999xa}-\cite{Altmannshofer:2008dz}, 
in $B _s \to \Phi( \to KK) l^+l^-$
\cite{Bobeth:2008ij},
and the simpler $B \to K l^+l^-$ decays \cite{Bobeth:2007dw}.
$\Lambda_b \to \Lambda$ decays offer further possibilities through polarization studies 
\cite{Aliev:2002tr}.

\subsection{$B \to K^*( \to K \pi)  l^+l^-$}

The full differential decay distribution (in $K^*$-zero-width approximation) can be written as
\beq d^4\Gamma =\frac{3}{8 \pi} J(q^2, \thl, \thK, \phi) dq^2 d
 \cos \thl d \cos \thK d\phi ,
 \eeq
  where
 \begin{eqnarray}
  J(q^2, \thl, \thK, \phi)& = &J_1^s \sin^2\thK + J_1^c \cos^2\thK
      + (J_2^s \sin^2\thK + J_2^c \cos^2\thK) \cos 2\thl
\nonumber \\       
    & +& J_3 \sin^2\thK \sin^2\thl \cos 2\phi 
      + J_4 \sin 2\thK \sin 2\thl \cos\phi 
      + J_5 \sin 2\thK \sin\thl \cos\phi
\nonumber \\      
    & + &J_6 \sin^2\thK \cos\thl 
      + J_7 \sin 2\thK \sin\thl \sin\phi
\nonumber \\ 
    & + &J_8 \sin 2\thK \sin 2\thl \sin\phi
      + J_9 \sin^2\thK \sin^2\thl \sin 2\phi , ~~~~~~J_i =J_i(q^2) .
  \label{eq:I:func}
\end{eqnarray}
Here, $\thl$ denotes the angle between the $l^-$ 
and the  $\bar B$ in the dilepton center-of-mass system (CMS)\footnote{Note that  the lepton angle  is also frequently  defined w.r.t. the $l^+$, and $\thl(\bar B l^-) = \pi-\thl (\bar B l^+)$. Here, $\bar B \equiv b \bar q$.},
$\thK$ is the angle between the $K$ and the $\bar B$ in the $K^*$-CMS and
$\phi$ is the angle between the normals of the $K \pi$ and the $l^+ l^-$ plane.
The angular distribution $d^4 \bar \Gamma$ of the
CP-conjugate decays is obtained after flipping the sign of the CP phases and by replacing $J_{1,2,3,4,7} \to \bar J_{1,2,3,4,7}$ and
$J_{5,6,8,9} \to - \bar J_{5,6,8,9}$.
The familiar $B \to K^* l^+l^-$ observables can be recovered as
$\Gamma \sim J_1- J_2/3$, $A_{\rm FB} \sim J_6$, $A_T^{(2)} \sim J_3$ \cite{Kruger:2005ep},
 $A_{\rm CP} \sim \Gamma -\bar \Gamma$ and the forward-backward CP asymmetry $A_{\rm FB}+ \bar A_{\rm FB} \sim A_{\rm FB}^{\rm CP}$ \cite{Buchalla:2000sk}.
 The angular analysis makes many more observables available.
Besides additional CP asymmetries \cite{Bobeth:2008ij}, discussed in Sec \ref{sec:CP},
further transverse asymmetries $A_T^{(3)}, A_T^{(4)}$ have been proposed
as "simple, clean, sensitive, precise" probes of the dipole couplings  $C_7^{(\prime)}$\cite{Egede:2008uy}. Some angular observables exhibit features known from $A_{\rm FB}$ in the sense that
a zero is present in the SM which can shift or go away in the presence of NP. 
Studies in many BSM models are performed  in Ref.~\cite{Altmannshofer:2008dz}.

\subsection{CP asymmetries \label{sec:CP}}

CP asymmetries in $b \to s$ transitions are doubly Cabibbo-suppressed 
$A_i \propto Im [ V_{ub} V_{us}^*/V_{tb} V_{ts}^*]  \sim 10^{-2}$
in the SM and any model where CP and flavor violation stems solely from the Yukawa matrices. 
Experimental investigations of  the $A_i$ are important tests of this paradigm.
{}From $d^4\Gamma$ and $d^4\bar \Gamma$ one can construct eight CP asymmetries $A_i \propto J_i -\bar J_i$  \cite{Kruger:1999xa}, sensitive to different Wilson coefficients   \cite{Bobeth:2008ij}.
The $A_{3,9}$ vanish in the SM by helicity conservation. They are sensitive to right-handed currents $C_i^\prime$.
The  $A_{3,9,(6)}$ can be extracted from a single-differential  distribution in $\phi (\theta_l)$.
The $A_{7,8,9}$ are T-odd and receive no suppression by small strong phases, such as those predicted by QCD factorization at low $q^2$\cite{Beneke:2004dp}.
The $A_{5,6,8,9}$ are CP-odd and can be extracted  without tagging from $\Gamma +\bar 
\Gamma$.
Both $A_7$ and $A_6$ are sensitive to $Z$-penguins ($\sim C_{10}$).
\begin{figure} 
\begin{center}
\includegraphics[width=.3\textwidth]{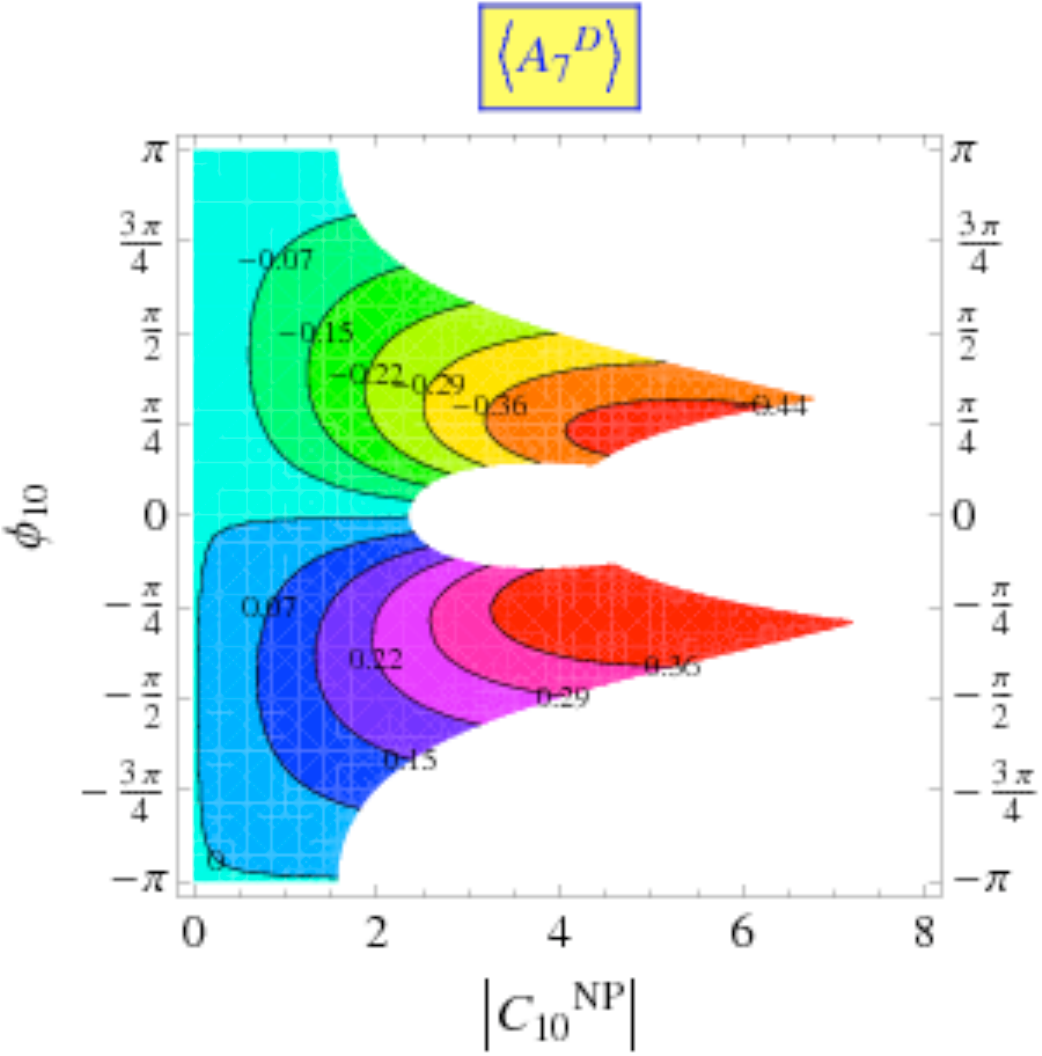} 
\includegraphics[width=.3\textwidth]{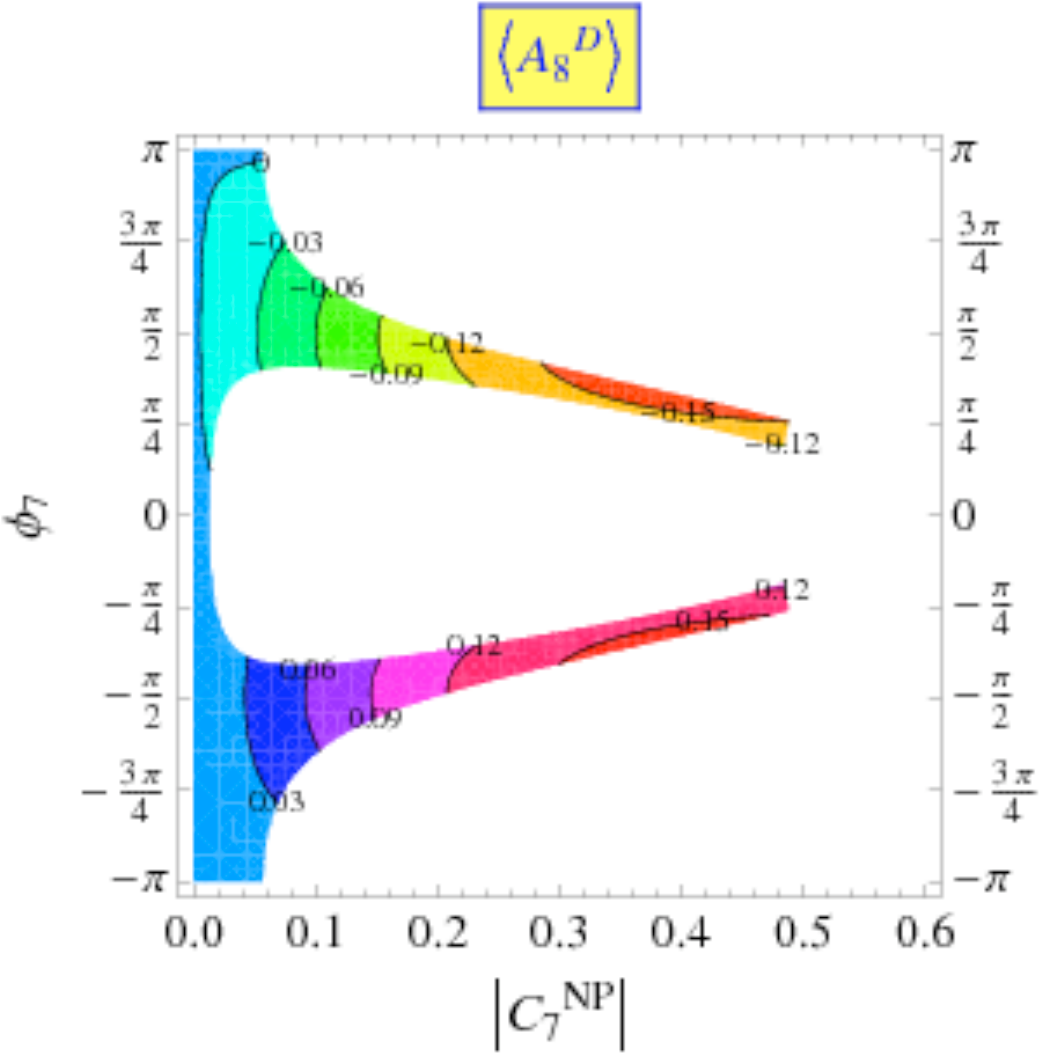} 
\includegraphics[width=.3\textwidth]{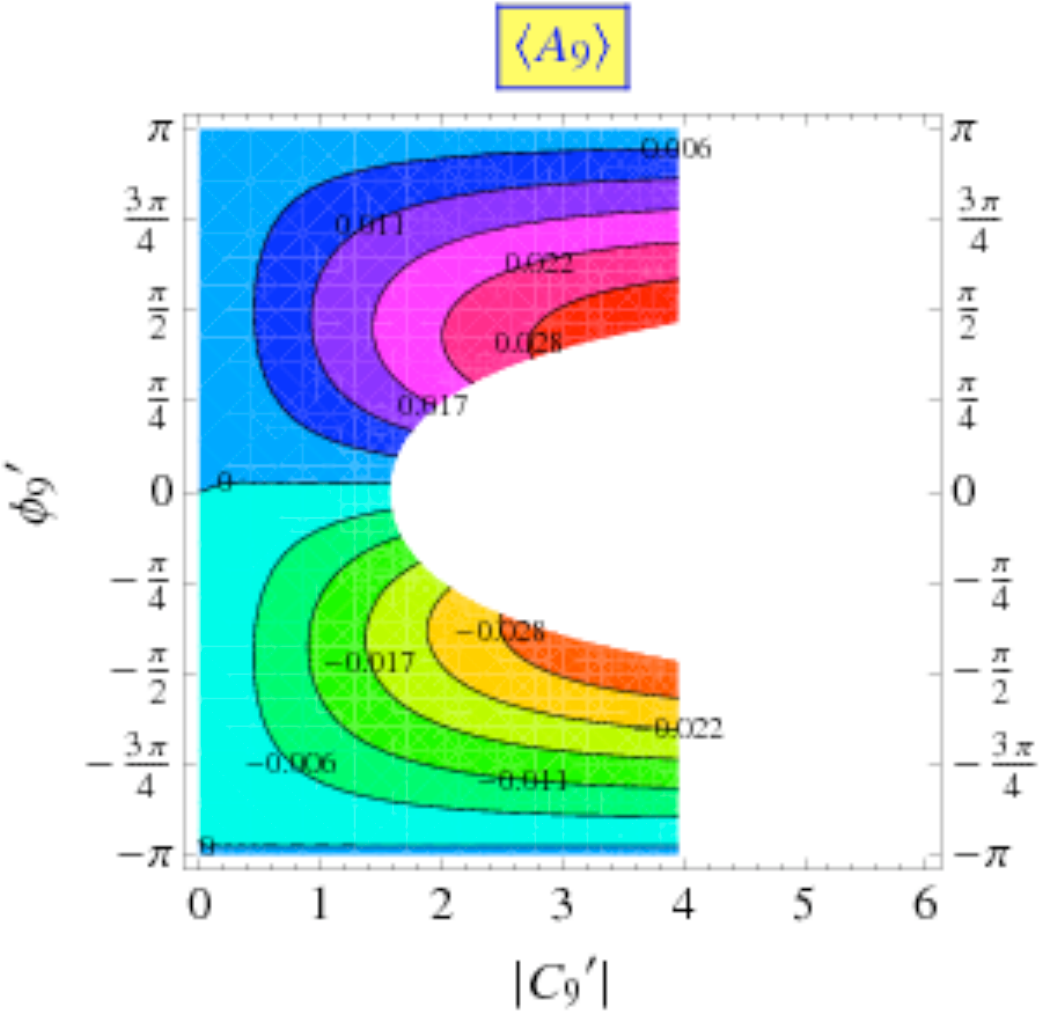} 
\includegraphics[width=.3\textwidth]{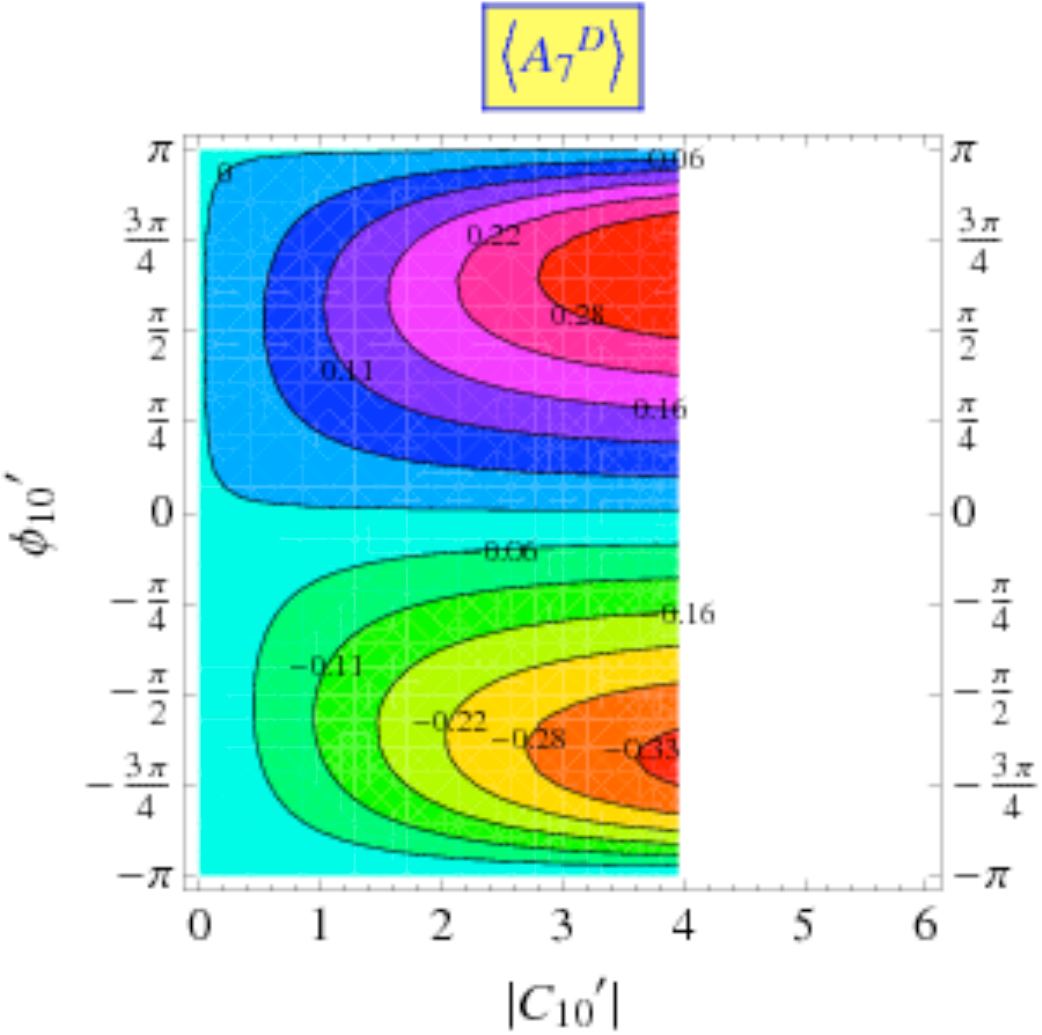} 
\includegraphics[width=.3\textwidth]{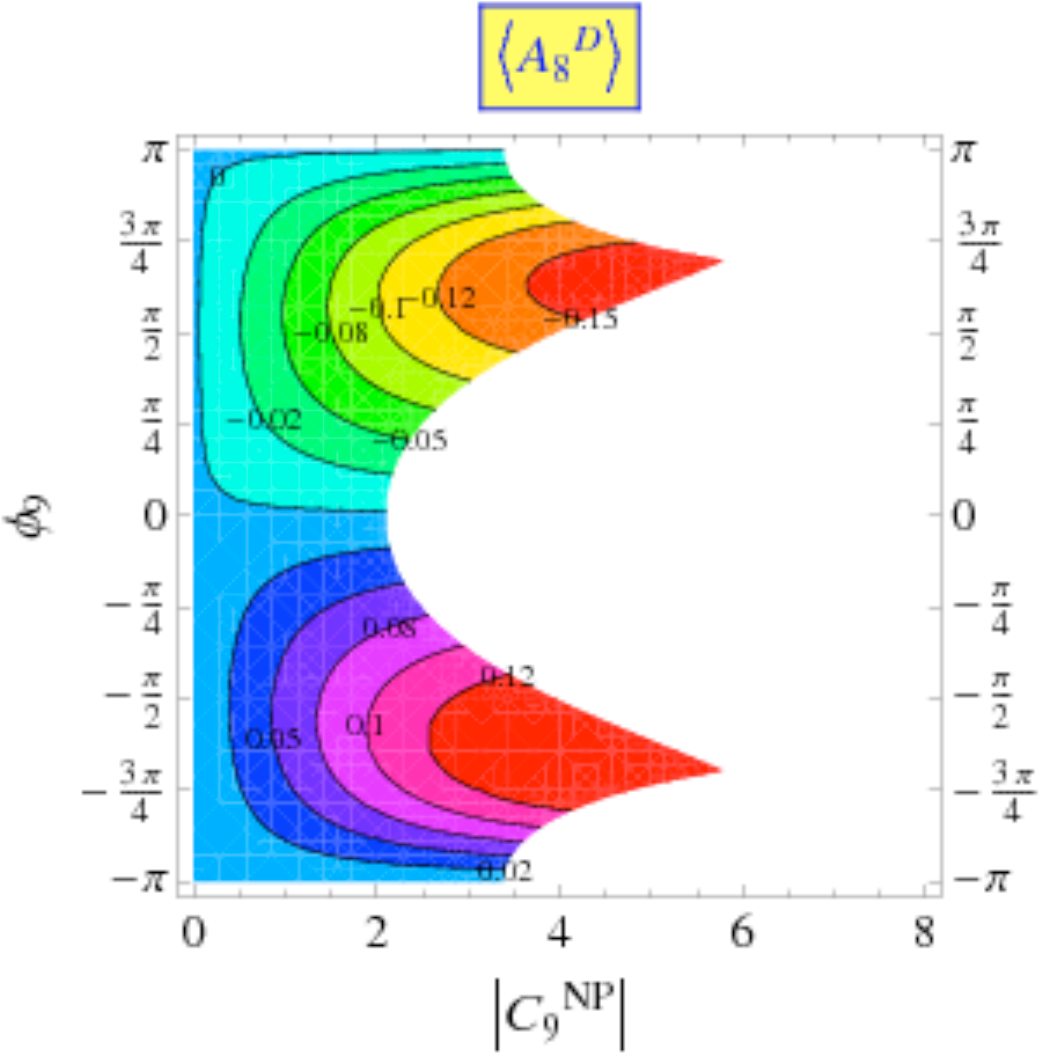} 
\includegraphics[width=.3\textwidth]{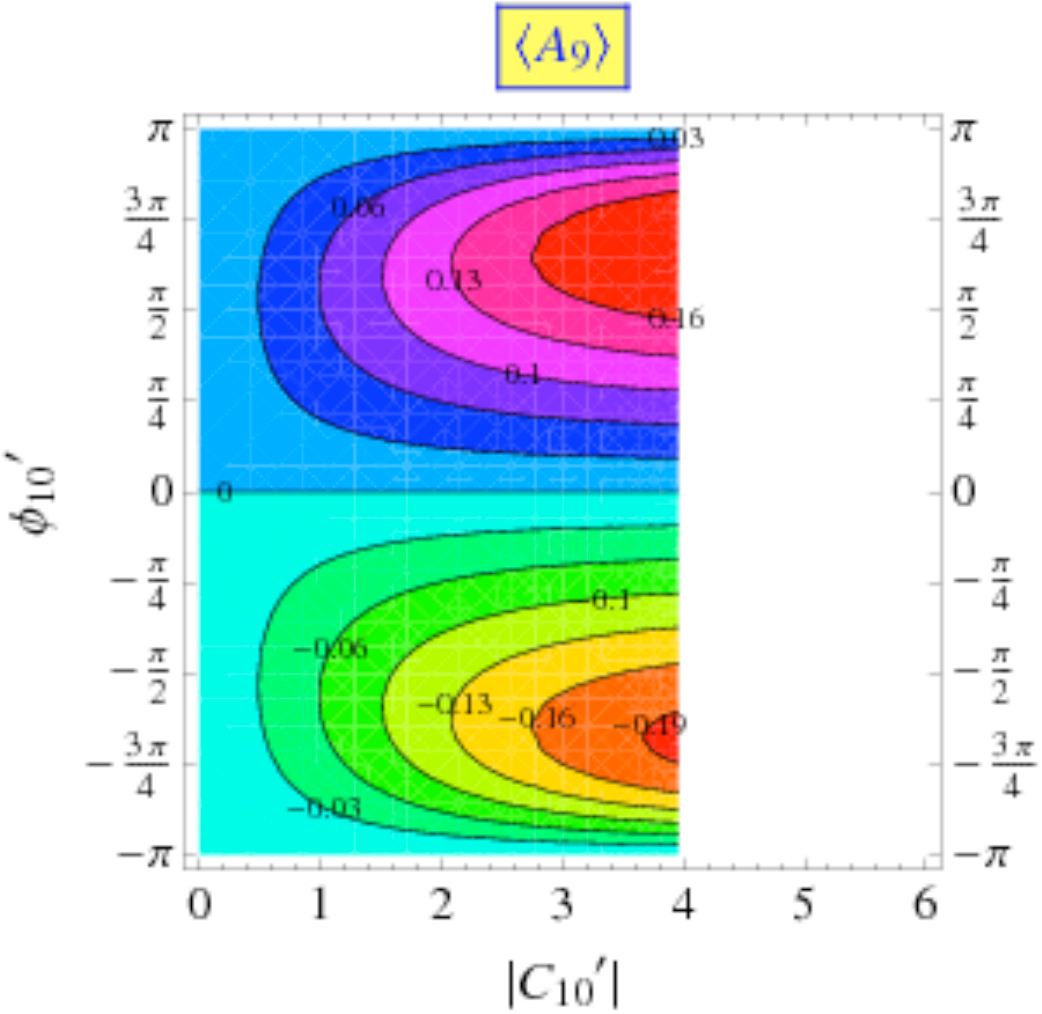} 
\includegraphics[width=.3\textwidth]{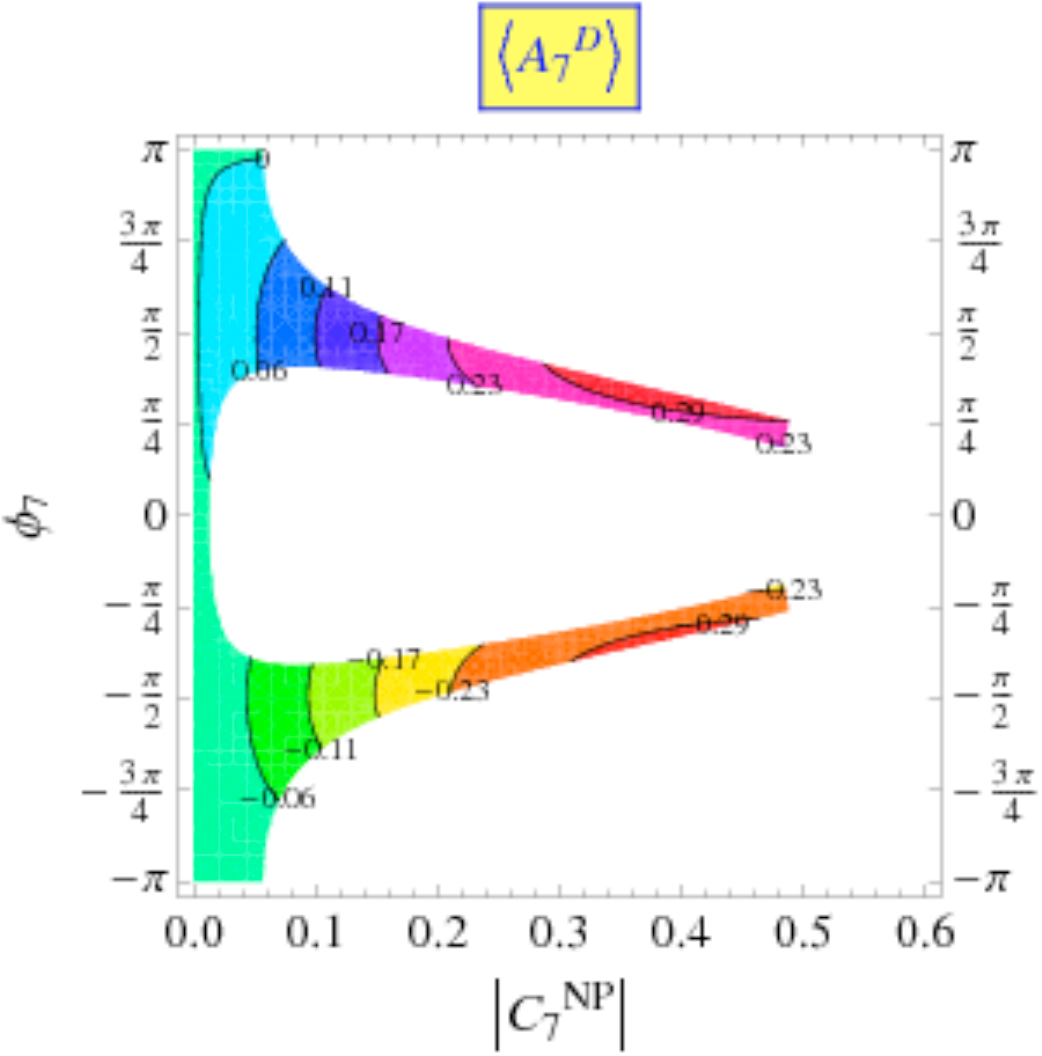} 
\includegraphics[width=.3\textwidth]{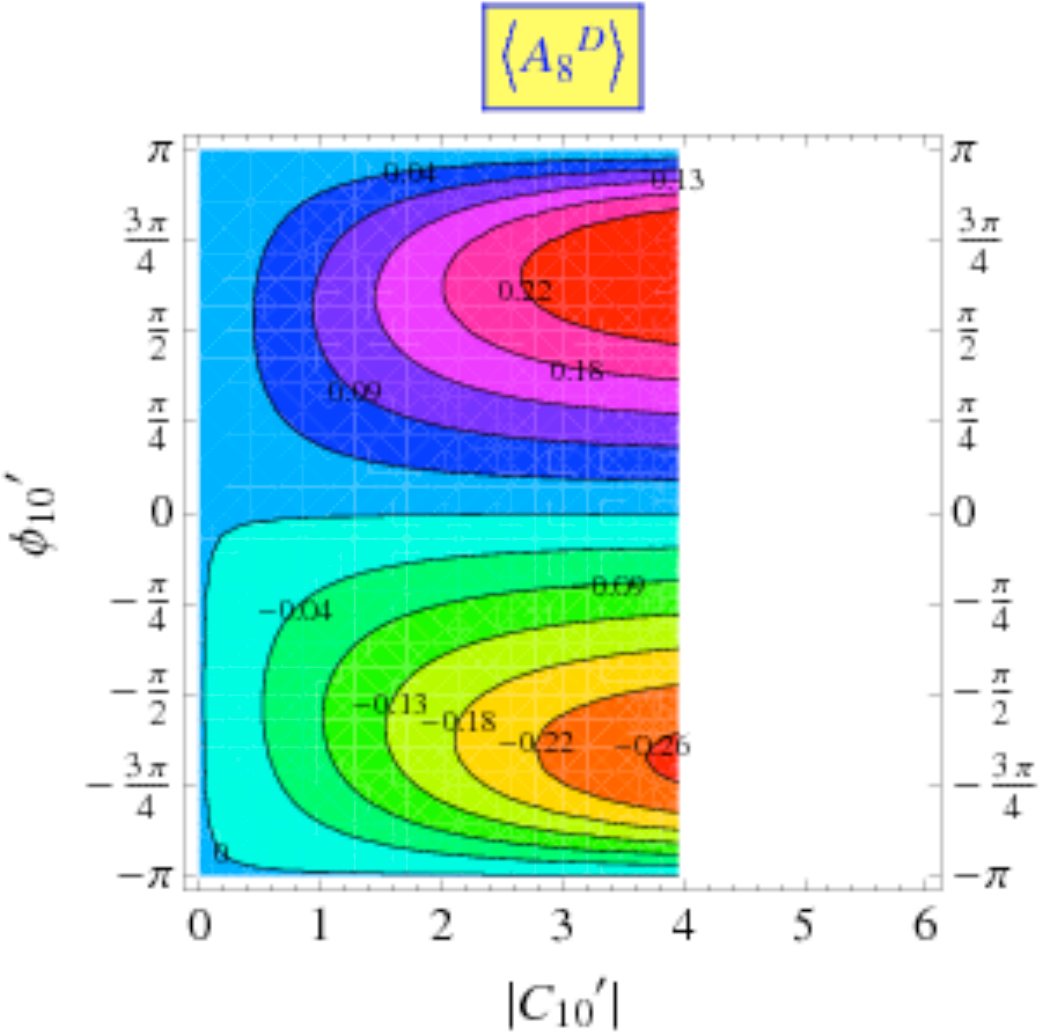} 
\includegraphics[width=.3\textwidth]{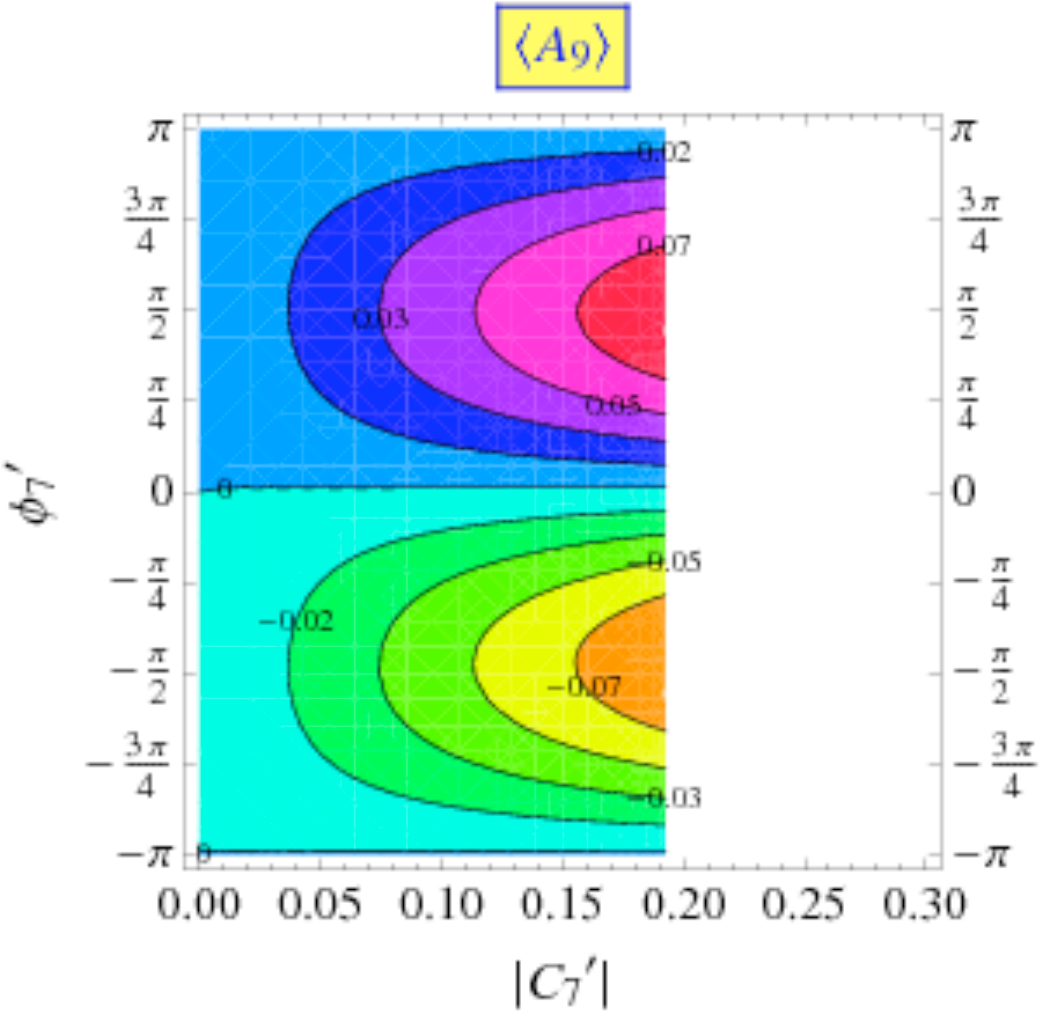} 
\end{center}
\caption{The (low $q^2$ integrated) T-odd CP asymmetries $A_{7,8}$ and $A_9$ 
depending on NP Wilson coefficients after
applying experimental constraints. Figures adopted from \cite{Bobeth:2008ij,Piranishvili:2008zz}.}
\label{fig:Todd} 
\end{figure} 
While T-even CP asymmetries $ \propto \sin \Delta_S \sin \Delta_W$ vanish for small strong phases $\Delta_S$, the T-odd asymmetries $ \propto \cos \Delta_S \sin \Delta_W$
exhibit maximal sensitivity to the  CP phases $\Delta_W$ in this limit. 
Contrary to the other $A_i$, the T-odd asymmetries $A_{7,8}$ and $A_9$ can be order one with NP (integrated over low $q^2$) \cite{Bobeth:2008ij}, see Fig.~\ref{fig:Todd}.
In each plot all other NP Wilson coefficients have been set to zero, and $B$ physics
constraints have been taken into account.

\subsection{$B_s \to \Phi( \to K K) l^+ l^-$}

The angular distributions in $B_s \to \Phi( \to K K) l^+ l^-$  allow to
study CP violation in interference between decay and $B_s$ mixing.
With the CP-odd asymmetries $A_{5,6,8,9}$ this is possible
without flavor-tagging \cite{Bobeth:2008ij}. (Unlike $\bar B_d, B_d \to K^*( \to K^\mp \pi^\pm) l^+ l^-$ or charged $B$ decays,
$B_s \to \Phi$ is not self-tagging).
The angular distribution of $B_s \to \Phi( \to K K) l^+ l^-$ is analogously defined as the one
of $B \to K^* (\to K \pi) l^+ l^-$ decays. Dominant differences between the decay amplitudes
originate from $SU(3)_F$ breaking. The biggest effects such as those from  form factors, decay constants are expected to cancel in the asymmetries. 
Significant differences between $B \to K^*$ and $B_s \to \Phi$ observables arise from the mixing properties. $B_s -\bar B_s$ mixing has a
substantial width difference $\Delta \Gamma_s/(2 \Gamma)  \sim {\cal{O}}(0.1)$ and
allows to measure time-integrated CP asymmetries, sensitive also to the $B_s$ mixing phase \cite{Bobeth:2008ij}. 

\subsection{$B \to K l^+l^-$}

The full angular distribution in $B \to K l^+l^-$ decays  can be written in terms of the decay rate
$\Gamma^l$, the  forward-backward asymmetry $A_{\rm FB}^l$ and a flat term, $F_H^l$, as
\cite{Bobeth:2007dw}
\begin{equation}
\frac{1}{\Gamma^l} \frac{d\Gamma^l}{d \cos \theta_l} =\frac{3}{4}(1-F_H^l) (1-\cos^2 \theta_l) +F_H
^l/2 +A_{\rm FB}^l \cos \theta_l .
\end{equation}
Here, the  dependence on the lepton species $l$ is kept to allow  for non-universal 
phenomena.
Other 
observables to probe such effects are $R_H ={\cal{B}}(B\to H \mu^+ \mu^-)/{\cal{B}}(B \to H e^+e^-)$,
$H=K^{(*)}, X_s$ \cite{Hiller:2003js}.
In the  SM the observables $F_H^l$, $A_{\rm FB}^l$ and $R_K-1$ are strongly  suppressed by the lepton mass and very small, and $\Gamma^l_{\rm SM} \propto  \sin^2 \theta_l$ \cite{Hiller:2003js,Bobeth:2007dw}.
Sizeable BSM effects are possible (here for low $q^2$)
\cite{Bobeth:2007dw}
\beq
|A_{\rm FB}^e| <13 \%, ~~~~~~~~~|A_{\rm FB}^\mu| < 15 \% ,~~~~~~~~~R_K-1={\cal{O}}(1) ,~~~~~~~~~
F_H^{e, \mu} < {\cal{O}}(0.5).
\eeq  
It follows that the forward-backward asymmetry of $B \to K l^+l^-$ cannot be neglected in a model-independent way. Correlations between the angular observables and $R_K$ and
further $b \to s l^+ l^-$ observables in several BSM scenarios 
are worked out in Ref.~\cite{Bobeth:2007dw}.

\section{Summary}

The coming years bring us large samples of flavor physics data from the LHC.
This way comes into reach a multitude of observables from exclusive $b \to s \mu^+ \mu^-$ 
processes, which allow to precisely map out the structure of the underlying physics. Decays specific to super flavor factories
into dielectrons $b \to s e^+ e^-$ and also
ditau and dineutrino modes provide further, complementary information. 
Quark flavor hierarchies in the SM predict $b \to d$ transitions to be suppressed with respect to $b \to s$ ones, in agreement with the observed values of $\Delta m_{d,s}$ and the 
$b \to (s,d) \gamma$ rates. It is open to test this  CKM-feature for semileptonic decays as well,
where first data on $B \to \pi l^+ l^-$ \cite{Wei:2008nv} have just become available.
My  favorite  semi-near term questions for $B$ decays with $l^+ l^-$ are:
$A_{\rm FB}$ (at low $q^2$),
${\cal{B}}(B_s \to \mu^+ \mu^-)$,
$R_{K, X_s} $ (improved), $F_H^l$,
${\cal{B}}(B_d \to \mu^+ \mu^-)/{\cal{B}}(B_s \to \mu^+ \mu^-)$, $A_{i (T)}$.

\acknowledgments
\vspace{-0.3cm}
This work is supported in
part by the Bundesministerium f\"ur Bildung und Forschung (BMBF)
and the German-Israeli-Foundation (G.I.F.).


\begin{thebibliography}{99}

\bibitem{Barberio:2008fa}
  E.~Barberio {\it et al.}  [Heavy Flavor Averaging Group],
  arXiv:0808.1297 [hep-ex]. \\Online http://www.slac.stanford.edu/xorg/hfag from September 2009.

\bibitem{beautymorello} Talk by M.J.Morello, this conference.

\bibitem{CDF9892} Public CDF note 9892.

\bibitem{:2008ju}
  B.~Aubert {\it et al.}  [BABAR Collab.],
  Phys.\ Rev.\  D {\bf 79} (2009) 031102
  [arXiv:0804.4412 [hep-ex]].
  
\bibitem{Aubert:2008ps}
  B.~Aubert {\it et al.}  [BABAR Collab.],
  Phys.\ Rev.\ Lett.\  {\bf 102}, 091803 (2009)
  [arXiv:0807.4119 [hep-ex]].
  
  
\bibitem{:2009zv}
  J.~T.~Wei {\it et al.}  [BELLE Collab.],
  Phys.\ Rev.\ Lett.\  {\bf 103}, 171801 (2009)
  [arXiv:0904.0770 [hep-ex]].


\bibitem{Hiller:2003js}
  G.~Hiller and F.~Kruger,
  Phys.\ Rev.\  D {\bf 69}, 074020 (2004)
  [arXiv:hep-ph/0310219].
  

\bibitem{Ali:1994bf}
  A.~Ali, G.~F.~Giudice and T.~Mannel,
  Z.\ Phys.\  C {\bf 67}, 417 (1995)
  [arXiv:hep-ph/9408213].
  
\bibitem{Bobeth:2007dw}
  C.~Bobeth, G.~Hiller and G.~Piranishvili,
  JHEP {\bf 0712} (2007) 040
  [arXiv:0709.4174 [hep-ph]].


\bibitem{Beneke:2004dp}
  M.~Beneke, T.~Feldmann and D.~Seidel,
  Eur.\ Phys.\ J.\  C {\bf 41} (2005) 173
  [arXiv:hep-ph/0412400].
  
\bibitem{Grinstein:2004vb}
  B.~Grinstein and D.~Pirjol,
  Phys.\ Rev.\  D {\bf 70} (2004) 114005
  [arXiv:hep-ph/0404250].



\bibitem{Bobeth:2008ij}
  C.~Bobeth, G.~Hiller and G.~Piranishvili,
  JHEP {\bf 0807} (2008) 106
  [arXiv:0805.2525 [hep-ph]].
  
\bibitem{Buchalla:2000sk}
  G.~Buchalla, G.~Hiller and G.~Isidori,
  Phys.\ Rev.\  D {\bf 63} (2000) 014015
  [arXiv:hep-ph/0006136].
  

\bibitem{Kruger:1999xa}
  F.~Kruger {\it et al.},
  Phys.\ Rev.\  D {\bf 61} (2000) 114028
  [Erratum-ibid.\  D {\bf 63} (2001) 019901]
  [arXiv:hep-ph/9907386].
  
\bibitem{Kruger:2005ep}
  F.~Kruger and J.~Matias,
  Phys.\ Rev.\  D {\bf 71} (2005) 094009
  [arXiv:hep-ph/0502060].
  
  
\bibitem{Egede:2008uy}
  U.~Egede {\it et al.},
  JHEP {\bf 0811} (2008) 032
  [arXiv:0807.2589 [hep-ph]].
  
\bibitem{Altmannshofer:2008dz}
  W.~Altmannshofer {\it et al.},
  JHEP {\bf 0901} (2009) 019
  [arXiv:0811.1214 [hep-ph]].
  
\bibitem{Aliev:2002tr}
  T.~M.~Aliev, A.~Ozpineci and M.~Savci,
  Phys.\ Rev.\  D {\bf 67}, 035007 (2003)
  [arXiv:hep-ph/0211447];
  G.~Hiller {et al.}, 
  Phys.\ Lett.\  B {\bf 649}, 152 (2007)
  [arXiv:hep-ph/0702191].
  
\bibitem{Piranishvili:2008zz}
  G.~Piranishvili, Ph.D.~thesis, Dortmund Tech.U., August 2008.

\bibitem{Wei:2008nv}
  J.~T.~Wei {\it et al.}  [Belle Collab.],
  Phys.\ Rev.\  D {\bf 78}, 011101 (2008)
  [arXiv:0804.3656 [hep-ex]].

\end{thebibliography}
\end{document}